Experimental Study on Shear Fatigue Behavior and Stiffness Performance of Warm Mix Asphalt by adding Synthetic Wax

**C. Petit** [a], **A. Millien** [a], **F. Canestrari** [b], **V. Pannunzio** [b], **A. Virgili** [b]

[a] *Université de Limoges, Laboratoire GEMH-GCD, Bd J. Derche, 19300 Egletons, France*
[b] *Politecnica Università delle Marche , P.zza Roma 22, 60121 Ancona, Italia.*

Corresponding Author : C. Petit e-mail : christophe.petit@unilim.fr , Phone : +33555934519

**Abstract**

Synthetic waxes produced by standard and registered processes may be used to manufacture Warm Mix Asphalt (WMA), which is a modified asphalt concrete produced, applied and compacted at temperatures below those typically required. This feature leads to environmental benefits, such as reduced energy consumption, gas and fume emissions, as well as to economic/operational advantages, such as lower production costs and greater hauling distances for extended construction seasons with tighter schedules. The present article serves to compare the mechanical performance of a WMA produced by adding synthetic wax with a traditional Hot Mix Asphalt (HMA) specimen, in terms of shear fatigue response and both complex and stiffness moduli. The experimental results and related modeling work demonstrate that adding synthetic wax into the WMA composition does not hinder either the destructive or non-destructive performance of an HMA, and this finding is corroborated by respectively measuring fatigue life and stiffness.

**1. Introduction**

Environmental considerations, focusing on the preservation and protection of natural and working environments, have become critical in several areas of civil engineering. In highway engineering for example, research and applications with environmentally-friendly materials and techniques have been expanding, and the development of Warm Mix Asphalt (WMA) is to be included in this category. The ability to generate asphalt mixtures that reduce production and application temperatures also yields a number of environmental benefits [1, 2], along with economic and operational advantages. Lower temperatures during asphalt production inevitably lead to reducing $CO_2/CO_2$ equivalent and fume emissions [3], which in turn guarantees better working conditions and less energy consumption during the production phase. The economic and operational benefits encompass: lower production costs, less fuel required for production, greater emissions control, potentially longer hauling distances, and extended construction schedules with tighter constraints [4].

WMA production proceeds according to several techniques. At present, two main groups of techniques can be distinguished: the first uses additives in order to modify asphalt fluidity, while the second group is based on binder foaming. The first group uses synthetic soluble organic substances (i.e. waxes), which are able to modify the rheological characteristics of asphalt binder by reducing its viscosity at a fixed temperature. The wax influences a temporary reversible change in the physical state, as characterized by melting and crystallization [5].

The second group of techniques combines preheated binder at 180°C with small amounts of cold water, which induces a binder foaming process. In this case, as the water comes into contact with hot bitumen, it turns to steam and thus favors a rapid expansion of bitumen in bubble form. This foam increases the specific surface of bitumen and reduces its viscosity, leading to a thorough coating of aggregates while lowering the operating temperatures of asphalt mixtures [6]. WMA produced using synthetic zeolite is to be included in this second group [7].

The synthetic waxes used to produce WMA can be manufactured by means of a Fisher-Tropsch process (FT), which is based on the coal gasification of a fine crystalline whose carbon chain ranges in length from $C_{45}$ to over $C_{100}$, with a melting point between 90° and 100°C. Synthetic waxes differ from macrocrystalline bituminous paraffin, which are naturally included in asphalt binder, featuring a carbon chain length in the $C_{25}$ to $C_{50}$ range and a melting point between 45° and 70°C. The longer carbon chains in the FT wax lead to a higher melting point; in contrast, the smaller structure reduces brittleness at low temperatures, as opposed to bitumen paraffin waxes [8]. Hence, synthetic waxes can be considered as a "binder flow improver"; as such, at temperatures above 120°-130°C, these waxes tend to lower the dynamic viscosity of asphalt binder, which then allows decreasing the mixing temperature while maintaining a complete aggregate coating. At temperatures below their melting point, synthetic waxes form a crystalline structure in the binder, which improves the asphalt mixture stiffness [9]. A traditional binder may be modified by adding synthetic waxes directly in the laboratory or asphalt mixing plant, at a rate ranging from 0.8% to 3% by weight of asphalt binder [10].

The Università Politecnica delle Marche (Italy), in cooperation with the Université de Limoges (France) and the University of Wisconsin-Madison (USA), has been involved in a research project on WMA with synthetic wax addition.

Pavements are subjected to fatigue while the surface layers (particularly overlays) undergo cyclic loading due to traffic (with shearing as the main source) [11-13]. This study focuses on the primary behavior of WMA used in surface layers, in comparison with HMA. The relevant research is also part of the aforementioned project and entails an experimental investigation of both destructive and non-destructive mechanical performance, including fatigue resistance in a shear configuration and both complex and stiffness moduli, for the purpose of drawing a WMA-HMA mechanical comparison.

## 2. Primary goals

The present study has been undertaken to compare the performance, in terms of shear fatigue behavior, compression complex modulus and indirect tensile stiffness modulus, of a WMA containing synthetic wax with a traditional HMA. The fatigue tests were conducted on parallelepiped asphalt concrete samples at two temperatures and one frequency, while the modulus tests were carried out on conventional cylindrical samples at three temperatures and several loading frequencies. The experimental fatigue results were then analyzed by considering two distinct failure criteria, so as to derive additional information on possible differences in fatigue response; the first criterion is simply based on a reduction in shear modulus, and the second is a mathematical model based on the damage growth trend, according to which the inflection point on the damage vs. number of fatigue cycles curve is assumed to be the failure point.

## 3. Laboratory experimental program: Materials and testing

### 3.1 *Materials and sample preparation*

Two types of asphalt concretes, WMA and HMA, were studied by use of a typical gradation curve (see Table 1) for binder course hot-mix asphalt concrete, classified as AC 16 bin 70/100, in accordance with the UNI EN 13108-1 Standard.

*Table 1: HMA and WMA aggregate gradation*

| Sieve [mm] | 32 | 16 | 10 | 4 | 2 | 0.5 | 0.25 | 0.063 |
|---|---|---|---|---|---|---|---|---|
| % passing by weight | 100 | 95 | 79 | 50 | 33 | 20 | 14 | 6 |

Limestone aggregates and two asphalt binders, with a binder content equal to 5.2% of the mix, were supplied to produce all of the asphalt mixture specimens tested. The HMA mixtures were prepared with a plain 70/100 pen binder; the WMA preparation introduced a 70/100 pen binder modified by adding synthetic wax of the Sasobit® type (FT) at a rate of 3% by mass of binder. This amount was chosen in order to assess the mechanical effects of adding the highest rate suggested by the manufacturer. The asphalt modification step was completed directly in the laboratory by adding the wax into the hot binder (at 150°C) and then mixing by means of a portable mixer for 30 minutes.

Concerning the mixing and compaction temperatures, the HMA samples were mixed at 160°C and compacted at 150°C, while WMA samples underwent mixing at 120°C and compaction at 110°C. The samples subjected to fatigue testing measured 50 mm x 70 mm x 125 mm and were obtained by cutting full asphalt concrete slabs (measuring 305 mm × 305 mm × 60 mm) that had been compacted by a roller compactor apparatus run at a preset pressure and number of passes. The samples undergoing stiffness characterization were cylindrical (Φ = 100 mm) with heights of 150 mm for the compression complex modulus and 60 mm for the indirect tensile stiffness modulus; they were obtained by using a Shear Gyratory Compactor (SGC), in accordance with UNI EN 12697-31. All HMA and WMA samples tested were measured at an (5 ± 0.7)% air void content to allow for a reliable comparison of mechanical performance.

3.2  *Testing program and instrumentation*

The complex modulus characterization steps were carried out in a compression state using a Dynamic Creep Apparatus, in accordance with the NCHRP 9-29: PT 01 procedure. The 150-mm high cylindrical samples were tested by setting a sinusoidal load shape under sweep frequency conditions ranging from 0.5 to 20 Hz. All tests were performed in a load control mode and three temperatures (0°, 10° and 20°C) were applied. Lastly, the stiffness modulus test was conducted in an indirect tensile configuration by introducing a Nottingham Asphalt Tester (NAT) apparatus, in accordance with UNI EN 12697-26. The 60-mm high cylindrical samples were tested with a rise time of 1.25 µsec (≈ 2 Hz) at 0°, 10° and 20°C under a strain control mode capped at 5 µm. For both test samples, a conditioning step ran for at least 6 hours before testing, and 5 replicates were tested for each considered mix.

The fatigue test was carried out using Mechanical Testing System (MTS). A new shear testing apparatus [14] was designed and applied within the framework of a joint research project sponsored by the GEMH Laboratory and France's Eurovia Research Center located in Mérignac. The Double Shear Test (DST) (i.e. double shear symmetric fatigue test), run in shear mode, has been designed to provide reliable tack coat or bituminous concrete material parameters relative to localized damage. The test set-up, which was integrated with a closed-loop servo hydraulic feedback system, uses a displacement or loading control mode.

This device is able to perform direct shear tests on asphalt concrete slabs with standard dimensions of 50 mm x 70 mm x 125 mm, into which four 10-mm high notches have been cut near the loaded central part (60-mm width), in order to generate and guide shear band localization (as shown in Figs. 1 and 2). A scale effect is witnessed for large-sized aggregates, and a Bazant Law is assumed to be available for such tests. Larger samples are complicated due to the high loading level; in the future therefore, it would be convenient to identify the Bazant Law for this device. A sinusoidal load shape at a 10-Hz frequency in a load control mode was introduced and two temperatures, 10° and 20°C, were investigated. These loading and temperature parameters are commonly used to simulate loading conditions in pavement base layers. In the present study, the stiffness is decreasing due to heating and fatigue, as is the case in standard fatigue studies on asphalt materials. For each mixture, a total of six samples were tested at 10°C and four at 20°C. Data measured on the evolution of shear modulus, phase angle and relative displacement vs. number of cycles/time were collected and then analyzed.

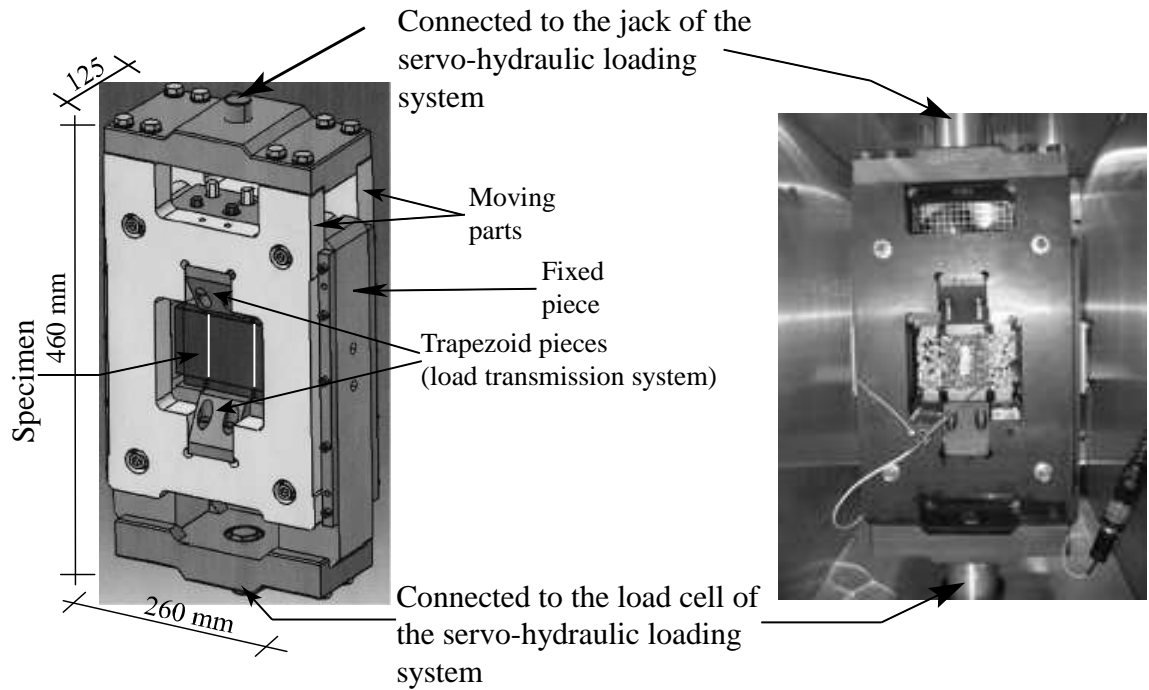

*Figure 1: Double Shear Test Device*

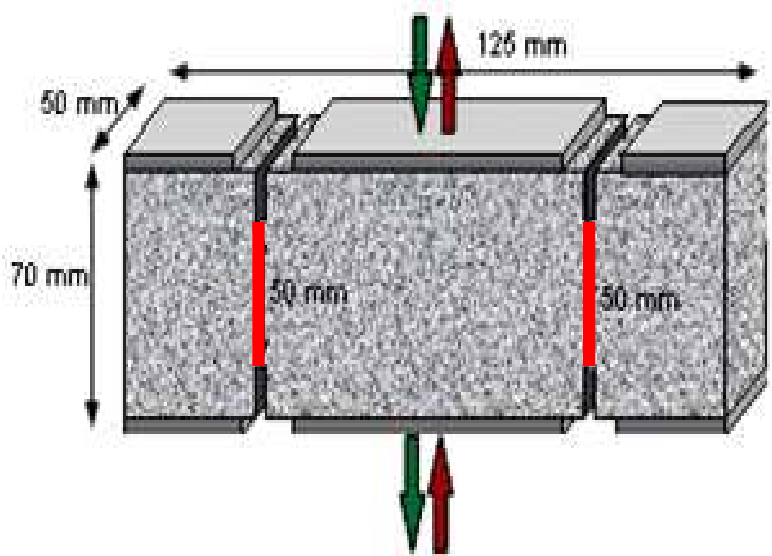

*Figure 2: Representation of the shear band (in red) between both notches*

## 4. Data generation: Modeling of measurement data

The fatigue response of asphalt mixes may be interpreted by considering various failure criteria [15-17]. In this work, fatigue test results have been derived by examining two failure criteria, called here the "50% reduction in modulus" and the "flex point model". The first criterion assumes that failure during the fatigue life occurs when the initial stiffness has decreased by 50%. In contrast, the flex point model is based on a mathematical analysis that considers fatigue as a gradual and continuous evolution of damage inside the asphalt concrete specimen. The beginning of a fatigue test is characterized by a decreasing rate of evolution in a selected fatigue parameter followed by a gradual increase up until failure. The evolution rate is ascribed a zero value at the flex (or inflection) point, which can thus be assumed as the point of failure.

This model, initially introduced by Virgili *et al.* [18], has been further developed using the Matlab software for application to DST fatigue results. This model is mainly based on the assumption that damage grows everywhere within the shear band (see Fig. 2). Damage can thus be described through a dimensionless scalar variable D, whose value ranges from 0 (undamaged material) to 1 (complete failure). In this study, it is assumed [13] that microdamage is localized within a band between opposite notches. The band length is denoted *c* and remains unchanged during the damage phase; moreover, *c* is decreasing as macrocracks are growing at the end of the test. According to such an approach, a band model with surface damage dissipation (instead of volumetric dissipation) has been considered. Consequently, a relatively uniform distribution of shear stress in the band may be assumed. The behavior of the joint is given by (1):

$$\tau = K_{s0}^* \cdot u \tag{1}$$

where τ is the peak of mean shear stress in the band, and u is the peak of relative band displacement, $K_{s0}^*$ is the undamaged complex shear stiffness. Then during the test, the complex shear stiffness is decreasing in the same time damage is increasing and displacement is increasing; only shear stresses stay constant.

With this approach, the effective shear modulus of a damaged material can be defined according to (2) as:

$$K_s^* = K_{s0}^* \cdot (1-D) \tag{2}$$

where $K_{s0}^*$ is the complex shear stiffness of undamaged material, as measured during the fatigue test in a shear configuration.

Damage evolution $\dot{D} = \frac{\partial D}{\partial t}$ can now be described by a power law, as written in Equation (3):

$$\dot{D} = \alpha \cdot D^{\beta} \tag{3}$$

where α and β are variable material parameters.

Starting with this equation, it is possible to obtain a relationship between the damage factor D and number of fatigue cycles N, as represented in Equations (4) and (5), where $R_k$ is a parameter that takes non-equidistant acquisition data into account:

$$D_k = \left[(1+R_k)\cdot (D_{k-1})^{1-\beta} - R_k \cdot (D_{k-2})^{1-\beta}\right]^{\frac{1}{1-\beta}} \tag{4}$$

where:

$$R_k = \frac{N_k - N_{k-1}}{N_{k-1} - N_{k-2}} \tag{5}$$

Parameter $\beta$ influences the damage law through material behavior. All values potentially assumed by parameter β may be divided into four main cases, but the only important case for the fatigue failure criterion is when β equals zero. In fact, β = 0 identifies the flex point where the damage evolution rate as a function of number of cycles reverses its trend [18, 19]. This unanimously determined point has been adopted as the failure criterion.

The proposed model applied to the fatigue test considers that β is a progressive function of the number of fatigue cycles *N*, as reported in (6):

$$\beta_k = \chi_7 + \chi_8 \cdot N_k + \chi_9 \cdot N_k^2 \tag{6}$$

where $\beta_k$ is the k[th] value of parameter β, corresponding to the k[th] fatigue cycle $n_k$, while $\chi_7$, $\chi_8$ and $\chi_9$ are regression parameters.

Regarding $K_s^*$, the following well-known sigmoidal function [20] was selected to describe shear modulus evolution from one fatigue cycle to the next:

$$\log K_s^* = \delta + \frac{\zeta}{1+\exp[\psi - \gamma \cdot (\log f + \log a_T)]} \tag{7}$$

Where δ the logarithm of the minimum modulus value, ζ the difference between logarithms of maximum and minimum modulus values, and ψ and γ are shape parameters. The correlation between the load frequencies and shift factor is described in (8):

$$\log f_{fict} - \log f = \log a_T \tag{8}$$

where $f$ is the measured load frequency, $f_{fict}$ the reduced frequency value and $a_T$ the shift factor, which has been determined herein by considering the classical Williams-Landel-Ferry (WLF) Equation [21].

The final equation applied can then be written as follows (9):

$$K_{sk}^* = 10^{\wedge}\left\{\delta + \frac{\zeta}{1+\exp\left[\psi - \gamma \cdot \left(\log f - \frac{C_1 \cdot (T_n - T_r)}{C_2 \cdot (T_n - T_r)}\right)\right]}\right\} \cdot \left\{1 - \left[(1+R_k) \cdot (D_{k-1})^{1-\beta_k} - R_k \cdot (D_{k-2})^{1-\beta_k}\right]^{\frac{1}{1-\beta_K}}\right\} \quad (9)$$

Equation (9) corresponds to equation (2) written for the k-th load cycle. Varying the value of the k index, equation (9) allows the construction of the effective shear modulus curve for the damaged material, corresponding to the damage D given by equation (4) for the same load cycle. Details related to the theoretical formulation of the analytical model introduced in equation (9) have been presented in a previous work [18].

In order to solve this equation, an additional parameter denoted $D_{in+1}$ was introduced. In imposing a damage value at the beginning of the test equal to 0, the damage factor value of $D_{in+1}$ is assumed to be an iterative parameter. Moreover, the model regression parameters (i.e. δ, ζ, ψ, γ, $C_1$, $C_2$, $\chi_7$, $\chi_8$, $\chi_9$ and $D_{in+1}$) can now be iteratively obtained by minimizing the sum of squared errors between experimental data and model values.

Experimental results for the complex modulus derived by use of the Dynamic Creep Apparatus were evaluated with the well-known master curve criterion. The sigmoidal model depicted in (7) was chosen to model the measurement data.

## 5. Results and comments

### 5.1 *Fatigue behavior*

Experimental fatigue data can be represented in terms of relative displacement peak ($v$) and dissipated energy (DE) vs. the number of cycles to failure.

Moreover, classical linear laws in bi-log scale, of the type $v = a(N_f)^b$ and $DE = c(N_f)^d$, have been selected as fatigue laws, in which $N_f$ is the number of cycles to failure, as defined by a 50% reduction in modulus and the flex point model, where $a$, $b$, $c$ and $d$ are material parameters. The fatigue line and related parameters are therefore able to describe the fatigue life behavior of a predefined asphalt mixture. Differences in initial shear modulus were verified at both test temperatures, with WMA samples always showing a slightly higher stiffness than the verified HMA samples. Such differences might depend on the crystallization of synthetic wax at temperatures below the melting point, a supposition that would ascribe an added stiffness to the asphalt mixture (as also verified in previous studies on WMAs [7, 9]).

The dissipated energy criterion is well known to interpretations conducted in asphalt concrete mechanics [22]; for this study, the proposed formula is introduced in Equation (10):

$$DE_n = \pi \cdot \tau \cdot u \cdot \sin\theta \qquad (10)$$

where $DE_n$ is the dissipated energy [in J/mm²] within the band corresponding to the n$^{th}$ cycle, $\tau$ the peak of shear stress, $v$ the peak of relative displacement, and $\theta$ the phase angle.

An attempt to omit the direct dependence of fatigue response on temperature has been based on combining the experimental data collected at both test temperatures with measured results fit at 10° and 20°C by means of a conventional power law. The first fatigue dataset was generated by considering a 50% reduction in modulus value as the failure criterion; Figures 3 and 4 show the final results obtained in terms of peak of relative displacement and dissipated energy of the undamaged material (25$^{th}$ cycle) vs. number of cycles to failure.

In examining Figure 3, several conclusions can be drawn. A linear law seems to be reliable for interpreting the measurement data collected at various test temperatures. The $R^2$ correlation parameters are high enough for both mixtures tested, as reported in Table 2. As for the HMA-WMA comparison, slight differences in fatigue responses can be observed. It thus seems that HMA and WMA fatigue responses depend on the initial deformation level applied, hence on the initial stiffness properties (as has been corroborated by a previous study [9]). This trend can also be remarked when simply considering the measured experimental data in Figure 3. By focusing on Table 2 however, it can be seen that fatigue law parameters "a" and "b" are very similar. In contrast, Figure 4 and Table 2 indicate that a linear law is able to fit fatigue results, in terms of dissipated energy vs. fatigue cycles to failure, at various temperatures with better reliability. In fact, correlation values in this case exceed 0.9 for both mixtures considered. HMA-WMA differences are more pronounced, especially when considering fatigue responses at low levels of dissipated energy, at which point WMA appear to behave better than HMA. For high dissipated energy levels, WMA and HMA fatigue laws are very similar. These slight differences in fatigue laws are more noticeable in Table 2, where the gaps in parameters "c" and "d" between mixes are greater than those verified with the shear relative displacement; these findings also depend on the different value ranges considered for DE. This initial data presentation reveals the possibility of representing fatigue laws for asphalt mixtures in two distinct ways when considering measurement data generated from different test temperatures. Lastly, in light of these initial results, it would seem that adding synthetic wax to produce WMA and the corresponding lower working temperatures do not undermine the shear fatigue response of a traditional HMA.

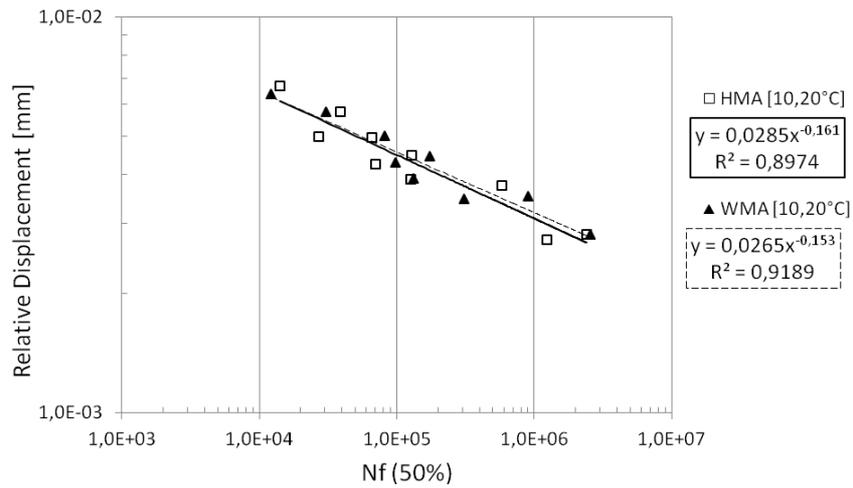

*Figure 3:* *Fatigue responses, as measured by relative displacement vs. cycles to failure with a 50% reduction in modulus*

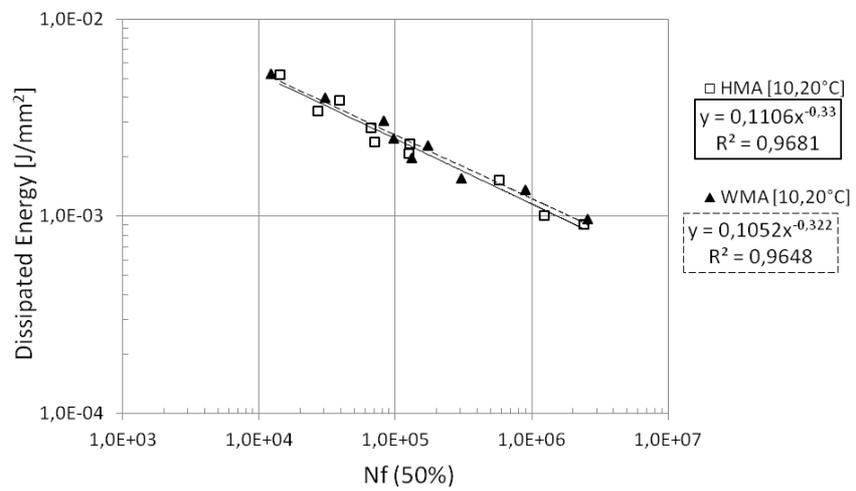

*Figure 4:* *Fatigue responses, as measured by dissipated energy vs. cycles to failure with a 50% reduction in modulus*

*Table 2: Fitting parameters for power fatigue laws (failure with a 50% reduction in modulus)*

| Mixture | $R^2$ | a | b | y-axis |
|---------|-------|---|---|--------|
| HMA | 0.8974 | 0.0285 | -0.161 | relative displacement |
| WMA | 0.9189 | 0.0265 | -0.153 | |
| Mixture | $R^2$ | c | d | y-axis |
| HMA | 0.9681 | 0.1106 | -0.330 | dissipated energy |
| WMA | 0.9648 | 0.1052 | -0.322 | |

Measurement data were then analyzed by applying the flex point model in order to define new fatigue failure points and compare the corresponding fatigue responses with those found above. As previously mentioned, the model was implemented using the Matlab software; Figures 5 through 7 offer an example of typical analytical results plotted for an individual fatigue test.

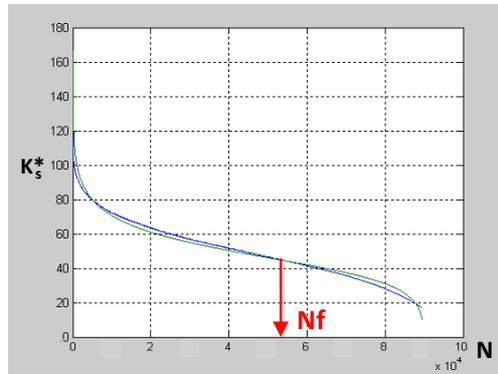

*Figure 5: Evolution in shear modulus ($K_s^*$) vs. number of cycles (N), with experimental data fitted by the model*

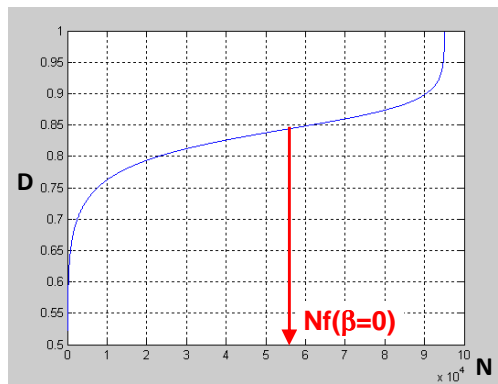

*Figure 6: Typical evolution of parameter D vs. number of cycles (N)*

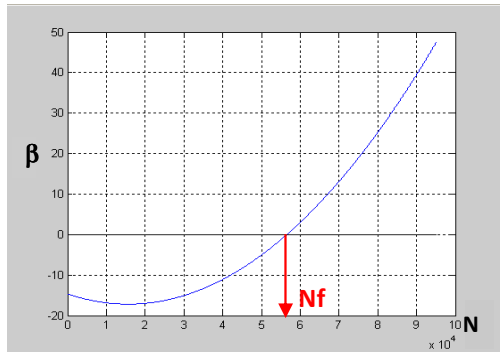

*Figure 7:* *Typical evolution of parameter β vs. number of cycles (N)*

Figure 7 displays a classical evolution in parameter β, which in the proposed model proves fundamental to determining the actual flex point, which has been proposed here as the fatigue failure point. When β equals 0, the corresponding related number of cycles is considered as the number to failure.

Figures 8 and 9 depict fatigue laws that consider the flex point model criterion.

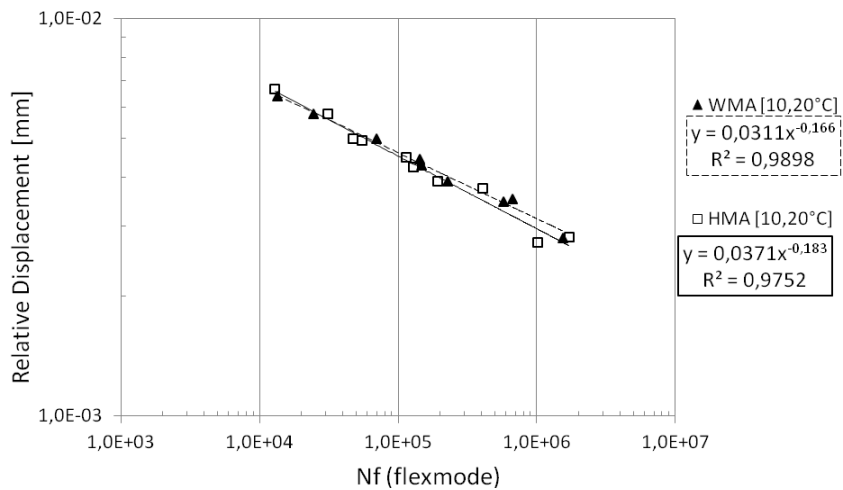

*Figure 8:* *Fatigue responses, viewed as relative displacement vs. cycles to failure with the flex point model*

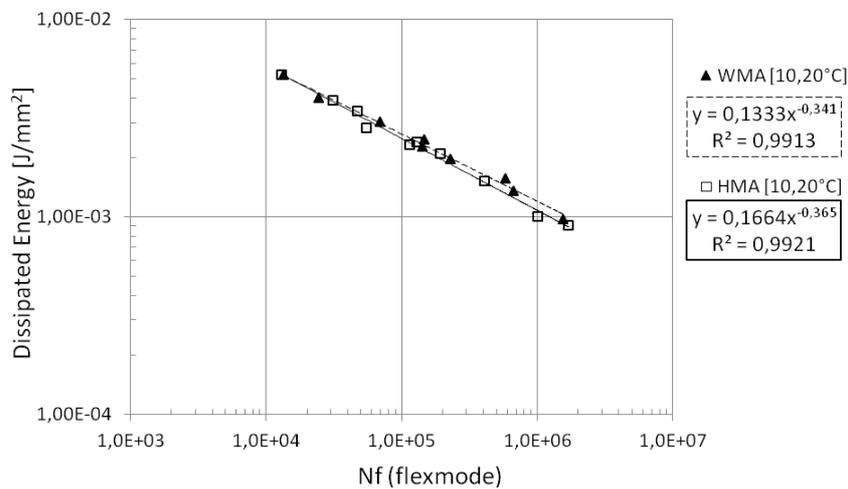
12

*Figure 9: Fatigue responses, viewed as dissipated energy vs. cycles to failure with the flex point model*

*Table 3: Fitting parameters for power fatigue laws (failure with the flex point model)*

| Mixture | $R^2$ | a | b | y-axis |
|---------|-------|---|---|--------|
| HMA | 0.9752 | 0.0371 | -0.183 | relative displacement |
| WMA | 0.9898 | 0.0311 | -0.166 | |
| Mixture | $R^2$ | c | d | y-axis |
| HMA | 0.9921 | 0.1664 | -0.365 | dissipated energy |
| WMA | 0.9908 | 0.1333 | -0.341 | |

In evaluating Figure 8, it can readily be observed that a power law applied with the flex point model offers great reliability when interpreting measured data collected at various test temperatures. Correlation parameters $R^2$ are greater than 0.9, which indicates a very good correlation between measured data and the classical linear fatigue law. Moreover, in comparing Tables 2 and 3 for the purpose of plotting in relative displacements, a good similarity in material parameters yet with different correlation values can be seen. This finding allows stating that a fatigue life representation is more reliable when applying the flex point model, even though parameters "a" and "b" are very similar for both failure criteria and both mixtures. By plotting results in terms of relative displacement and then changing the failure criterion, the material parameters of both mixtures therefore maintain very similar and comparable values. As displayed in Figures 3, 4, 8 and 9 differences in response can be detected for low initial loading levels. In comparing HMA and WMA responses, parameters "c" and "d" slightly differ; in particular, the gap in parameter "c" is characterized by a factor of roughly 1.25. Hence, the differences in material responses are distinct and seem to depend on the level of dissipated energy. As can be verified in figure 3, 4, 8 and 9, WMA appears to perform better at low energy levels, while a similar performance can be observed for HMA at higher energy levels.

Lastly, Tables 4 and 5 below list detailed results of the flex point model application, along with a number of comparisons between the two failure criteria proposed herein.

*Table 4: Detailed modeling results for samples tested at 10°C*

| Sample code | Test temp. [°C] | $R^2$ | $N_f$(50%) | $N_f$(flexmode) | % $K_{s0}^*$ [flexmod] |
|---|---|---|---|---|---|
| HMA 2-7 | 10 | 0.984 | 14,235 | 12,914 | 54.6 |
| HMA 2-5 | 10 | 0.973 | 38,744 | 31,203 | 58.5 |
| HMA 1-2 | 10 | 0.960 | 66,153 | 54,692 | 59.0 |
| HMA 2-3 | 10 | 0.956 | 127,752 | 113,799 | 54.8 |
| HMA 1-5 | 10 | 0.972 | 579,821 | 409,743 | 63.5 |
| HMA 1-4 | 10 | 0.921 | 2,413,967 | 1,704,769 | 61.6 |
| WMA 2-5 | 10 | 0.965 | 12,245 | 13,371 | 46.8 |
| WMA 1-5 | 10 | 0.970 | 30,565 | 24,337 | 57.9 |
| WMA 2-6 | 10 | 0.955 | 82,078 | 69,473 | 56.7 |
| WMA 1-1 | 10 | 0.980 | 174,355 | 142,733 | 58.5 |
| WMA 3-8 | 10 | 0.983 | 895,815 | 665,587 | 58.4 |
| WMA 1-2 | 10 | 0.978 | 2,568,015 | 1,555,693 | 65.5 |

*Table 5: Detailed modeling results for samples tested at 20°C*

| Sample code | Test temp. [°C] | $R^2$ | $N_f(50\%)$ | $N_f$(flexmode) | % $K_{s0}^*$ flexmode] |
|---|---|---|---|---|---|
| HMA 1-1 | 20 | 0.988 | 26,930 | 47,264 | 37.9 |
| HMA 1-3 | 20 | 0.991 | 70,444 | 128,681 | 39.6 |
| HMA 2-6 | 20 | 0.982 | 125,991 | 191,919 | 40.5 |
| HMA 1-7 | 20 | 0.996 | 1,235,019 | 1,008,732 | 43.1 |
| WMA 3-4 | 20 | 0.987 | 97,651 | 147,451 | 40.8 |
| WMA 3-5 | 20 | 0.989 | 132,674 | 228,553 | 39.9 |
| WMA 3-1 | 20 | 0.977 | 306,786 | 577,281 | 35.6 |

From these two tables, several findings can be drawn. The correlations between the model proposed in this work and experimental results are very strong across all individual test data processed, as seen by the $R^2$ values (in excess of 0.9). This outcome allows concluding that the proposed model seems to be characterized by a great reliability when interpreting the shear fatigue results of asphalt concretes at different test temperatures. The values labeled as $N_f(50\%)$ and $N_f$(flexmode) are the number of cycles to failure determined, respectively, by a 50% reduction in modulus and by the flex point model, while % $K_{s0}^*$ [flexmode] is the ratio in percent between the modulus referred to in $N_f$, flex point, and the initial modulus value. A comparison between the two failure criteria employed now becomes possible. The temperature effects on the mixture can be evaluated by comparing the number of fatigue cycles to failure with the corresponding reduction in modulus value. Let's remark that the flex point model identifies various failure points besides those determined by a 50% reduction in modulus. At 10°C, the proposed model identifies fatigue failure before that identified by the 50% modulus reduction, with the direct consequence being that failure occurs whenever the reduction in stiffness is less than 50%, except in one case (WMA 2-5). The opposite arises for data recorded at 20°C, for which the modeled failure points are located after that expected by the 50% modulus reduction, with the direct consequence of failure occurring for a stiffness reduction of more than 50%. The results listed in Tables 4 and 5 thus highlight in detail the differences in fatigue failure between the two adopted criteria. Given that the 50% modulus reduction is a failure criterion based solely on a pre-established reduction, hence on a conventional test point, the proposed model reveals itself to be more highly correlated with the entire material response during testing.

## 5.2 Stiffness characterization

Stiffness characterization involves both the compression complex modulus (E*), as evaluated with respect to the master curve by the sigmoidal model introduced above, and the indirect tensile stiffness modulus (ITSM).

Figure 10 shows the various mixture responses through the measurement data and related master curves.

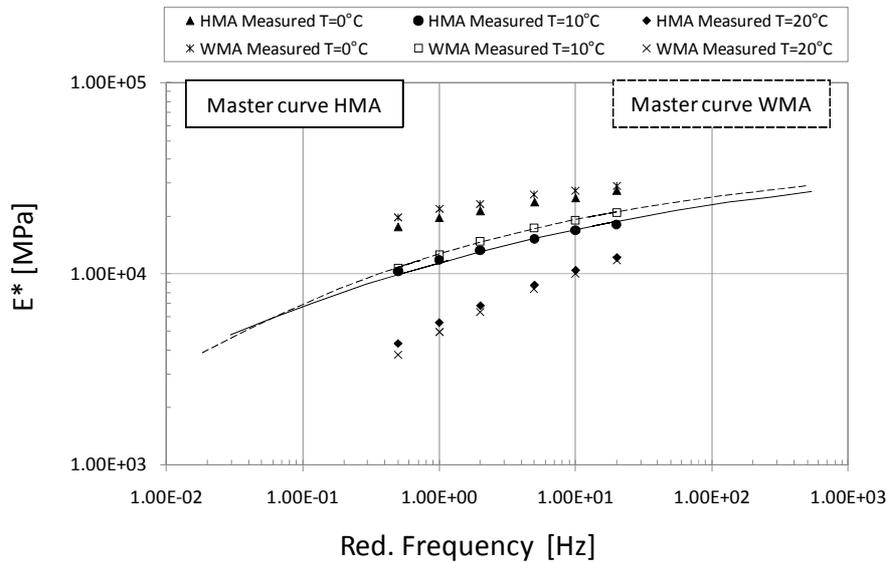

*Figure 10: Complex modulus: HMA-WMA measurement data and master curves*

As seen in Figure 10, the complex modulus responses of both asphalt mixtures are correlated with the reduced frequency and hence to the test temperatures being considered. This dependence is noticeable for both the measurement data and the final master curves derived using 10°C as the reference temperature. The master curves describe how an asphalt mixture behaves in terms of complex modulus vs. frequency, in omitting the direct dependence on temperature. The extra stiffness of WMA over a wide range of reduced frequencies above 1 Hz can be observed, thus indicating a mechanical response very similar to what was identified during other experimental studies [23]. However, WMA seems to lose stiffness at frequencies lower than 1 Hz, i.e. where WMA values lie closer to HMA. Around 0.1 Hz, an intersection is found between master curves, all of which exhibit practically the same response; moreover, a trend towards stiffness gap reduction is noticed for reduced frequency values above 100 Hz. The temperature sensitivity of mixtures is apparent when examining individual measurement data: at frequencies of less than 1 Hz, HMA shows higher modulus values than WMA, whose behavior can be recorded for an individual experimental dataset at 20°C.

Table 6 lists all sigmoidal model parameters, whose values (except for δ) of the mixtures tested remain quite similar. In addition, $R^2$ values are high, and this finding exhibits a strong model results - experimental data correlation.

**Table 6:** *Sigmoidal model parameters for master curves ($T_{ref.} = 10°C$)*

| Mixture code | δ | ζ | Ψ | γ | $R^2$ |
|---|---|---|---|---|---|
| HMA | 0.0937 | 4.577 | -1.858 | 0.420 | 0.985 |
| WMA | 0.3939 | 4.187 | -2.033 | 0.530 | 0.978 |

**Table 7:** *ITSM: HMA-WMA measurement data*

| Mixture | HMA | | | WMA | | |
|---|---|---|---|---|---|---|
| Temperature [°C] | 0 | 10 | 20 | 0 | 10 | 20 |
| ITSM [MPa] | 17571 | 11215 | 5694 | 18947 | 10766 | 4753 |
| Std. dev. [MPa] | 1346 | 1053 | 680 | 1566 | 469 | 318 |

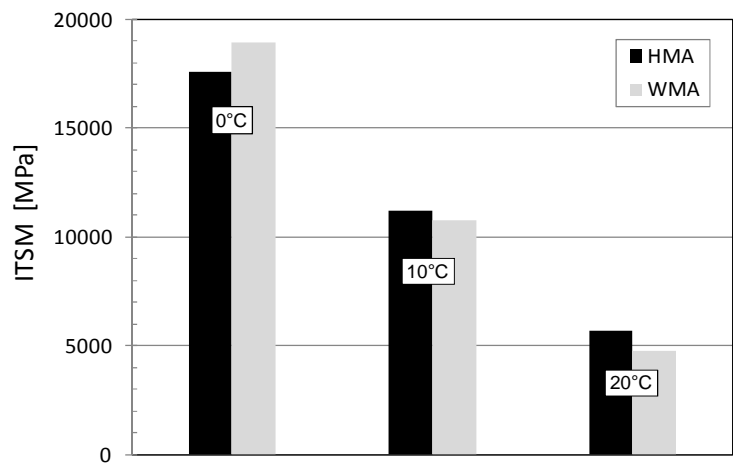

*Figure 11: ITSM: HMA-WMA measurement data*

Figure 11 and table 7 present the experimental stiffness data recorded using an indirect tensile configuration. The experimental data displayed in Figure 11 suggest that HMA and WMA stiffnesses are strictly correlated with test temperature. A decrease in stiffness occurs in both mixtures as temperature increases. When comparing Figures 10 and 11 and under similar test conditions, it is obvious that the applied load configuration plays a remarkable role in the stiffness response of both mixtures, as revealed in the mechanical performance of an asphalt mixture being higher in compression than in indirect tension. The additional WMA stiffness only appears at 0°C, as the warm samples tend to increase in stiffness around the 10% level. The opposite occurs at other temperatures, where WMA exhibits a decrease in modulus around 4% at 10°C, and then around 15% at 20°C. As also observed in the compression configuration at low frequencies, WMA stiffness properties show visible temperature sensitivity. Because all samples tested, whether HMA or WMA, were very close to 5% air void content, the measured additional WMA stiffness at low temperatures can only be correlated with the presence of synthetic wax. Once crystallized, the wax serves to add stiffness to any asphalt binder or mixture, and this finding can be verified during testing at predefined loading frequencies and temperatures.

## 6. Conclusion

This research program has studied the mechanical properties, in terms of shear fatigue and stiffness, of both WMA by adding synthetic wax and HMA for the purpose of drawing comparisons with one another. Samples of both mixes, with the same gradation and asphalt content, a different shape and a very similar air void content, were tested under various test conditions regarding test configuration, loading frequency and test temperature. Fatigue results were analyzed using two failure criteria: a 50% reduction in modulus, and the flex point model. The final responses were depicted relative to shear relative displacement and dissipated energy vs. number of cycles to failure. Moreover, stiffness properties were assessed with respect to compression complex modulus master curves and indirect tensile stiffness modulus values.

In light of the range of results obtained and reported above, the following conclusions can be drawn:

- Good overall reliability from the use of various failure criteria, as well as different representations and the classical power fatigue laws have all been verified by analyzing fatigue data obtained with two test temperatures. The proposed model, defined as the "flex point model", has proven to be an effective mathematical tool for interpreting shear fatigue data. An array of result representations, including relative displacement and dissipated energy vs. number of cycles to failure, can yield different model-experimental result correlations, thus leading to differing reliable responses for the given failure criterion; however, both failure criteria may be utilized to interpret fatigue results at various test temperatures;
- The proposed fatigue representations, shown as WMA fatigue responses relative to both experimental data points and fatigue regression parameters, are very similar to those identified for traditional HMA, with differences depending on the criterion used and, more generally, on the relative displacement or dissipated energy level considered. Some noteworthy WMA-HMA differences have been found for the dissipated energy vs. cycles to failure parameter by applying the flex point model, for which the WMA seems to behave better. Moreover, the change in fatigue criteria implies a change in fatigue law parameters. The addition of synthetic wax does not therefore seem to degrade the fatigue behavior of an asphalt mixture;

- Stiffness characterization revealed differences relative to both loading frequency and temperature. WMA exhibited an added stiffness compared to HMA, especially at higher reduced frequencies and/or lower temperatures, as also verified in subsequent studies. This additional stiffness stems from the synthetic wax, which adds "stability" to an asphalt binder and consequently to the asphalt mixture due to its crystallized form.

## 7. References

[1] Stroup-Gardinier M. and C. Lange. *Characterization of Asphalt Odors and Emissions*. Proceedings of 9[th] International Conference on Asphalt Pavements. Copenhagen, Denmark, 2002.

[2] Van Wlik R. *The environmental management act, energy efficiency in Dutch asphalt industry*. 3[rd] Euroasphalt & Eurobitume Congress, Vienna, 2004.

[3] Hugener M., Emmenegger L., Mattrel P. *Emissions of tar-containing binders: a laboratory study*. Journal of Environmental Science and Health, Part A (2007), 42.

[4] The Asphalt Pavement Association of Oregon. *Warm Mix Asphalt Shows Promise for Cost Reduction, Environmental Benefit*. Centerline. The Asphalt Pavement Association of Oregon, Salem, OR, 2003.

[5] Edwards Y., Isacsson, U. 2005a. *Wax in Bitumen Part 1 – Classifications and general aspects*. Road Materials and Pavement Design (International Journal) Volume 6 – Issue 3/2005.

[6] Olard, F., Antoine, J.-P., Héritier, B., Romanier, A., Martineau, Y. *LEA® (Low Energy Asphalt): A new generation of asphalt mixture*. Proceedings of the Advanced Characterization of Pavement and Soil Engineering Materials, Athens, Greece, 2007.

[7] Bocci M., Grilli A., Pannunzio V., Riviera P.P. *Mechanical characterization and influence of water on WMA by adding synthetic zeolite*. Int. J. for Pavement Engineering & Asphalt Technology (PEAT), Vol. 10, Issue 2, 2009.

[8] Edwards Y., Isacsson, U. 2005b. *Wax in Bitumen Part 2 – Characterization and effects*. Road Materials and Pavement Design (International Journal) Volume 6 – Issue 4/2005.

[9] Cardone F., Pannunzio V., Virgili A., Barbati S. *An evaluation of use of synthetic waxes in warm mix asphalt*. 7th International RILEM Symposium on advanced testing and characterization of bituminous materials, Rhodes, 2009.

[10] Hurley G.C. and Prowell B.D. *Evaluation of Sasobit® for use in warm mix asphalt*. NCAT Report 05-06. National Center for Asphalt Technology, Auburn, AL, 2006.

[11] Bouhas F., Millien A., Petit C. *Finite element modeling of fatigue shear tests: Contribution to pavement lifetime design under traffic loading. Pavement cracking: Mechanisms, modeling, detection, testing and case histories*. 6th RILEM International Conference on Cracking in Pavements; Chicago, 2008, pp. 649-659.

[12] Su K., Sun L., Hachiya Y. and Maekawa R., *Analysis of shear stress in asphalt pavements under actual measured tire-pavement contact pressure*, Proceedings of the 6th ICPT, Sapporo, Japan, 2008, pp. 11-18

[13] Petit C., Laveissière D., Millien A., *Modelling of Reflective Cracking in Pavements: Fatigue under shear stresses,* Third Int. Symp. On 3D Finite Elements for Pavement Analysis, Design and Research, Amsterdam (The Netherlands), 2002, pp. 111-124.

[14] Diakhaté M, Millien A, Petit C, Phelipot-Mardelé A, Pouteau B. *Experimental investigation of tack coat fatigue performance: Towards an improved lifetime assessment of pavement structure interfaces.* Construction and Building Material 2011; 25(2):1123-33.

[15] Shatnawi S. R. *Fatigue performance of asphalt concrete mixes using a new repetitive direct tension test.* California Department of Transportation, 1994.

[16] Rowe G. M. and Bouldin M. G. *Fatigue damage analysis in asphalt concrete mixtures using the dissipated energy approach.* Can. J. Civ. Eng. 33, 2006.

[17] Ghuzlan A. K. and Carpenter S. H. *Improved Techniques to evaluate the fatigue resistance of asphaltic mixtures.* 2$^{nd}$ Eurasphalt & Eurobitume Congress, Barcelona, 2000.

[18] Virgili A., Partl M.N., Grilli A., Santagata F.A. *Damage model for environmental conditioned fatigue test with CAST.* Fatigue Fract Eng Mater Struct 2008, 31:967-979.

[19] Partl, M.N., Pasquini, E., Canestrari, F., Virgili, A. *Analysis of water and thermal sensitivity of open graded asphalt rubber mixtures.* Int. J. of Construction and Building Materials, Volume 24, Issue 3, 2010.

[20] Pellinen T. K. and Witczak M. *Stress Dependent Master Curve Construction for Dynamic Complex Modulus.* Annual Meeting of the Association Asphalt Paving Technologists, AAPT, Colorado, 2002.

[21] Williams, M. L., Landel, R. F., Ferry, J. D. *The temperature dependence of relaxation mechanisms in amorphous polymers and other glass-forming liquids*, Journal of the American Chemical Society, Vol. 7, 1955, p. 3701-3707.

[22] Ghuzlan K. A. and Carpenter S. H. *Fatigue damage analysis in asphalt concrete mixtures using the dissipated energy approach.* Can. J. Civ. Eng. 33(7), 2006.

[23] Silva H.M.R.D., Oliveira J.R.M., Pereira A.A.P. *Assessment of the performance of warm mix asphalts in road pavements.* II Int. Conf. Environmentally Friendly Roads ENVIROAD 2009, Warsaw, 2009.